# The tightly bound nuclei in the liquid drop model

N R Sree Harsha
South West Nuclear Hub,
University of Bristol, UK
*Email: sn16091@my.bristol.ac.uk*

**Abstract**

In this paper, we shall maximise the binding energy per nucleon function in the semi–empirical mass formula of the liquid drop model of the atomic nuclei to analytically prove that the mean binding energy per nucleon curve has local extrema at $A \approx 58.6960$, $Z \approx 26.3908$ and at $A \approx 62.0178$, $Z \approx 27.7506$. The Lagrange method of multipliers is used to arrive at these results, while we have let the values of $A$ and $Z$ take continuous fractional values. The shell model that shows why $^{62}$Ni is the most tightly bound nucleus is outlined. A brief account on stellar nucleosynthesis is presented to show why $^{56}$Fe is more abundant than $^{62}$Ni and $^{58}$Fe. We believe that the analytical proof presented in this paper can be a useful tool to the instructors to introduce the nucleus with the highest mean binding energy per nucleon.

## I.    Introduction

It is often incorrectly stated in many physics and astronomy textbooks that $^{56}$Fe is the most strongly bound nucleus [1–8]. However, Fewell and Shurtleff *et al.* have pointed out that both $^{58}$Fe and $^{62}$Ni are more strongly bound than $^{56}$Fe, with $^{62}$Ni having the highest mean binding energy per nucleon [9, 10]. In fact, Fewell has shown numerically that if the atomic number ($Z$), mass number ($A$) and the number of neutrons ($N = A - Z$) are allowed to take fractional values, the most tightly bound nucleus has $A \approx 58.3$, $Z \approx 26.6$ and a shift in binding energy curve favours $^{62}$Ni as the most tightly bound nucleus [9]. The calculations for the mean binding energy are often done numerically, using least-square fitting procedures on binding energy data of atomic nuclei, and many authors often state, without proof, that $A \approx 60$ represents the region of the highest mean binding energy per nucleon [11–13]. The recently determined mean binding energy per nucleon according to the experimentally available atomic masses for the six most tightly bound nuclei are shown in the table 1 [14].

**Table 1:** The experimentally determined mean binding energy per nucleon for the six most tightly bound nuclei. The data are taken from the AME2016 atomic mass evaluations [14].

| Nuclide | AME2016 *BE/A* (keV/*A*) |
|---|---|
| $^{62}$Ni | 8794.533±0.007 |
| $^{58}$Fe | 8792.250±0.006 |
| $^{56}$Fe | 8790.354±0.005 |
| $^{60}$Ni | 8780.774±0.006 |
| $^{54}$Cr | 8777.955±0.007 |
| $^{64}$Ni | 8777.461±0.007 |

The graph of these binding energy per nucleon values for these nuclei is plotted against $A$ in figure 1 and it can be seen that the two local maxima points are found near $A \approx 58$ and $A \approx 62$. The reason why a particular combination of ($N$, $Z$), for each value of $A$ in figure 1, has a high value of binding energy is explained in the next section.

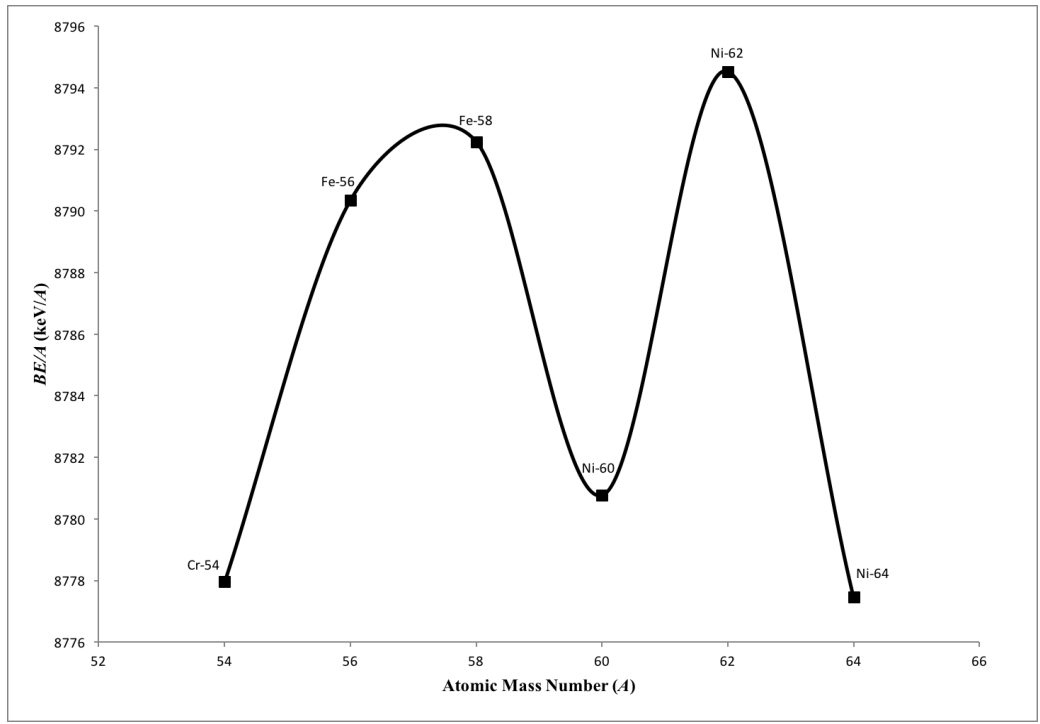

**Figure 1:** Mean binding energies per nucleon of the six most tightly bound nuclei with atomic mass numbers in the region $54 \leq A \leq 64$. They are all even–even nuclei, since owing to the parity term in the binding energy formula they have higher mean binding energy per nucleon compared to odd–even or odd–odd nuclei. The smooth solid line between the data points is a free–hand interpolation to guide the eye. The uncertainties are all smaller than the size of symbols used to represent the data points.

In this paper, we propose an analytical way, using the Lagrange method of multipliers, to find the nuclei with highest mean binding energy per nucleon. The analytical method is based on maximising the binding energy function in the semi-empirical mass formula (SEMF) of the liquid drop model of atomic nuclei. The form of the binding energy in the SEMF considered in this paper is described in section II. The Lagrange method to find out the local extrema of binding energy function is described in section III. Even though the liquid drop model of nuclei does not account for all the nuclear characteristics, the gross features of figure 1 are preserved and we should expect to find extrema around $A \approx 58$ and $A \approx 62$. The shell effects that favour $^{62}$Ni to be the most tightly bound nucleus are then presented. The relative high abundance of $^{56}$Fe compared to $^{62}$Ni is briefly described in section IV. The conclusions and the advantages of teaching the analytical proof are described in section V.

## II. The form of binding energy formula

The semi-empirical mass formulae (SEMF) have always been at the heart of our understanding of several properties of the atomic nuclei. The oldest and the simplest form of binding energy in SEMF, which is often referred to as Bethe-Weizsäcker (*BW*) formula [15], is

$$BE^{BW} = a_v A - a_s A^{2/3} - a_c \frac{Z(Z-1)}{A^{1/3}} - a_a \frac{(A-2Z)^2}{A} \qquad (1)$$

The coefficients $a_v$, $a_s$, $a_c$ and $a_a$ in equation (1) represent the volume, surface, Coulomb and asymmetry terms, respectively. It has been shown that these four terms of *BW* formula describe, with good accuracy, the various properties of nuclides such as fission, fusion, alpha–decay barrier potential energies [16, 17]. However, various suggestions have been made to add additional terms to *BW* formula to further improve the results. The various terms that are added are the Wigner term, pairing term, Coulomb exchange term, surface symmetry term etc. and these terms are described well in [18]. In this paper, we shall only consider addition of the pairing term to equation (1), since this term becomes significant while considering the nuclei with even number of protons and neutrons (often referred to as the even–even nuclei). Hence, the form of the binding energy we shall consider in this paper is

$$BE(A,Z) = BE^{BW} + BE_{pair} = a_v A - a_s A^{2/3} - a_c \frac{Z(Z-1)}{A^{1/3}} - a_a \frac{(A-2Z)^2}{A} + BE_{pair} \qquad (2)$$

The pairing term $BE_{pair}$ represents the effect of spin coupling between neutrons and protons in the nucleus and can be expressed as [18]

$$BE_{pair} = a_p \frac{\left[(-1)^{(A-Z)} + (-1)^Z\right]}{2\sqrt{A}} \qquad (3)$$

Here, $a_p$ is often referred to as the pairing coefficient. The latest comprehensive database of masses and binding energies of various nuclides, published in 2016, is the atomic mass evaluation, referred to as AME2016 [14]. The previous versions of the atomic mass evaluations were published in 2012 and 2003 and are referred to as AME2012 [19] and AME2003 [20], respectively. The coefficients of the Bethe-Weizsäcker formula based on the AME2016 database have not yet been determined. However, Royer and Subercaze [21] have determined the coefficients of the binding energy, based on 2027 nuclei with *N, Z* ≥ 8 using AME2012 database, assuming its form as shown in equation (4).

$$BE^{RS} = a_v^{RS}(1 - k_v I^2)A - a_s^{RS}(1 - k_s I^2)A^{2/3} - a_k^{RS}(1 - k_k I^2)A^{1/3} - \frac{3}{5}\frac{e^2 Z^2}{R_0}$$
$$+ f_p \frac{Z^2}{A} + BE_{pair} - BE_{shell} - BE_{Wigner} \qquad (4)$$

Here *I*, defined as $(A - 2Z)/A$, is called the relative neutron excess. The first term of equation (4) represents the volume term and asymmetry term of *BW* formula.

The second and third term represents surface and curvature energies. The fourth term represents the decrease of binding energy due to repulsion between protons in the nucleus. The fifth term represents the proton form-factor correction. The pairing term, correction due to the shell effects and the Wigner term are represented by $BE_{pair}$, $BE_{shell}$ and $BE_{Wigner}$ respectively. We rewrite equation (4) to separate the terms of the *BW* formula and the additional terms explicitly.

$$BE^{RS} = \left\{ a_v^{RS} A - a_s^{RS} A^{2/3} - \frac{3}{5}\frac{e^2 Z^2}{R_0} - a_v^{RS} k_v I^2 A + BE_{pair} \right\} - a_s^{RS} k_s I^2 A^{2/3} \\ - a_k^{RS}\left(1 - k_k I^2\right) A^{1/3} + f_p \frac{Z^2}{A} - BE_{shell} - BE_{Wigner} \right\} \quad (5)$$

The terms in the curly brackets are the terms in the *BW* formula with the addition of the pairing term. The question that we can now ask is can we, in order to convert $BE^{RS}$ to the binding energy *BE (A, Z)* shown in equation (2), let the values of $k_s$, $a_k$, $f_p$, $BE_{shell}$ and $BE_{Wigner}$ tend to zero?

Recently Kirson [18] has pointed out that the different terms in the binding energy formula, given by equation (5), are mutually dependent on each other using error matrix and studying correlation of various terms. For instance, he found that symmetry coefficient is equal to 22.5 MeV in the simple *BW* formula, but it increases to 31.5 MeV when all the various terms are included in the binding energy formula, as given by equation (5). He also noted that the pairing and shell correction terms are largely independent of the other terms. If we let the values of a few coefficients in the binding energy formula shown in equation (5) tend to zero, the values of other coefficients have to be changed to reproduce the experimental masses of the nuclides. Hence, in our notation, we can write that as $a_v \ne a_v^{RS}$, $a_s \ne a_s^{RS}$ $a_c \ne a_c^{RS}$ and $a_a \ne a_a^{RS}$ or, equivalently, we can express these conditions mathematically[1] as

$$BE(A,Z) \ne \lim_{\substack{k_s \to 0 \\ a_k^{RS} \to 0 \\ f_p \to 0}} \lim_{\substack{BE_{shell} \to 0 \\ BE_{Wigner} \to 0}} BE^{RS} \quad (6)$$

Therefore, while Royer and Subercaze have determined the coefficients of binding energy formula shown in equation (5) using the AME2012 database, we cannot use those coefficients in our *BE* formula shown in equation (2). We instead use the coefficients of the binding energy function determined by Royer and Gautier using AME2003 database [22]. They have determined that the binding energy function, that best fits experimental masses of 1522 nuclei with *Z, N > 7*, takes the following form

$$BE^{RG} = 15.7827 A - 17.9042 A^{2/3} - 0.72404 \frac{Z(Z-1)}{A^{1/3}} \\ - 23.7193 I^2 A + BE_{pair} - BE_{shell} \right\} \quad (7)$$

---

[1] The binding energy formulae can be thought of as rational functions with an introduction of a suitable variable (such as $k^6 = A$) and hence are continuous at every point in their domains. If the binding energy functions are considered to be functions of *A* and *Z*, their domains are represented as $A, Z \in (1, \infty)$. This continuity is also the reason we were able to draw a solid smooth line connecting the data points in figure 1.

As we have pointed out earlier, the shell correction term is independent of other terms in the binding energy formula and we shall, for the sake of simplicity, ignore it in our analysis. Hence, the binding energy formula that we shall be considering is represented mathematically as

$$BE(A,Z) = \lim_{BE_{shell} \to 0} BE^{RG} \tag{8}$$

Hence, we have the following binding energy formula, which we shall be considering for all the analyses presented in this paper.

$$BE(A,Z) = a_v A - a_s A^{2/3} - a_c \frac{Z(Z-1)}{A^{1/3}} - a_a I^2 A + a_p \frac{\left[(-1)^{(A-Z)} + (-1)^Z\right]}{2\sqrt{A}} \tag{9}$$

The values of the coefficients in equation (9) can be determined by comparing it with equations (7) and they are $a_v$ = 15.7827 MeV, $a_s$ = 17.9042 MeV, $a_c$ = 0.72404 MeV, $a_a$ = 23.7193 MeV and $a_p$ = 11 MeV.

The binding energy per nucleon of the six most tightly bound nuclei according to various atomic mass evaluations and theoretical calculations using equation (9) are shown in table 2 for comparison. It should be noted that the binding energies provided by different atomic mass evaluations only differ in their last two significant digits and we can use AME2003 database without terribly affecting the final results.

**Table 2:** The comparison of experimentally determined binding energy per nucleon of the six most tightly bound nuclei, as given in the various databases, with the theoretical calculations.

| Nuclide | AME2016 BE/A (keV/A) | AME2012 BE/A (keV/A) | AME2003 BE/A (keV/A) | Theoretical calculation (keV/A) |
|---|---|---|---|---|
| $^{62}$Ni | 8794.533±0.007 | 8794.546±0.008 | 8794.549±0.010 | 8828.78 |
| $^{58}$Fe | 8792.250±0.006 | 8792.239±0.008 | 8792.221±0.012 | 8832.18 |
| $^{56}$Fe | 8790.354±0.005 | 8790.342±0.008 | 8790.323±0.012 | 8811.51 |
| $^{60}$Ni | 8780.774±0.006 | 8780.764±0.008 | 8780.757±0.010 | 8797.24 |
| $^{54}$Cr | 8777.955±0.007 | 8777.935±0.011 | 8777.914±0.014 | 8822.59 |
| $^{64}$Ni | 8777.461±0.007 | 8777.454±0.009 | 8777.467±0.010 | 8819.33 |

The graph of the theoretical calculations of the binding energy per nucleon and the values taken from AME2003 are shown in figure 2. It can be seen there are a few differences between the theoretically predicted values and experimentally determined values. The binding energies determined from equation (9) are higher than their experimentally determined counterparts. This is because we have only considered five terms in our binding energy formula. The binding energy per nucleon of $^{58}$Fe is larger than $^{62}$Ni because we have neglected the shell correction effects in our binding energy formula. $^{62}$Ni has 28 protons and they form a closed shell and hence, $^{62}$Ni is expected to have higher mean binding energy per nucleon than $^{58}$Fe.

It should also be noted that the theoretical calculation of binding energy per nucleon for $^{54}$Cr is greater than $^{56}$Fe because the binding energy formula, even with the inclusion of all the terms, is only an approximate empirical formula. It cannot reproduce the exact values of binding energies of the nuclides. Nevertheless, for our analysis, it is sufficient to note that the theoretical calculations show two local extrema at $^{58}$Fe and $^{62}$Ni and we should expect that the Lagrange method, described in the section III, predicts local minima around $A \approx 62$ and $A \approx 58$.

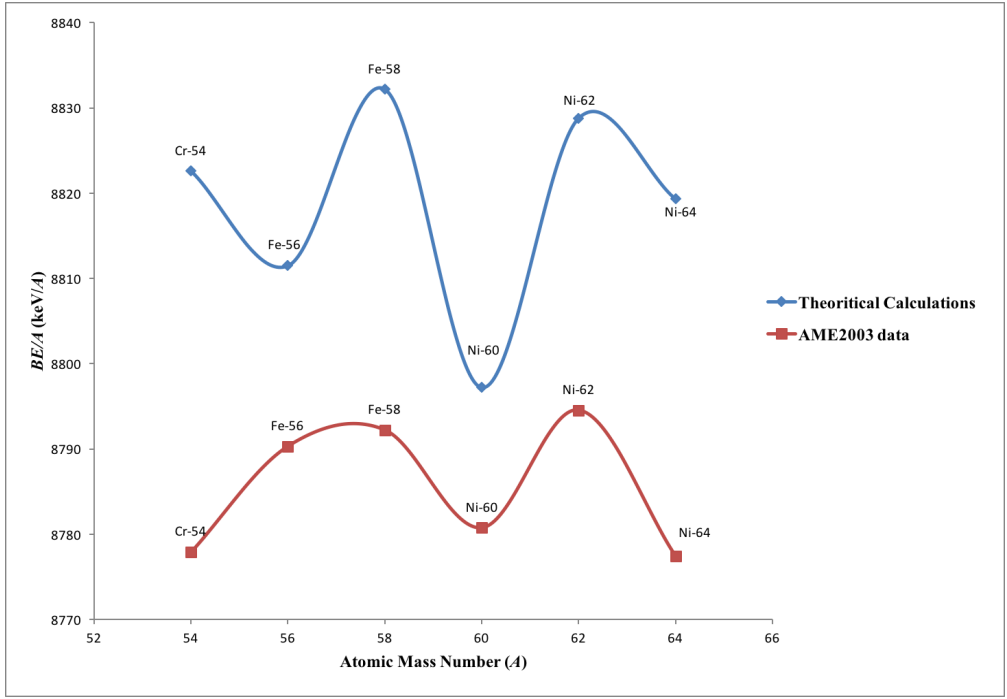

**Figure 2:** The experimentally determined binding energy per nucleon of six most tightly bound nuclei are plotted against mass number $A$ in red, while the theoretically calculated values are plotted in blue. The solid smooth lines in both the cases are free-hand interpolations to guide the eye. The uncertainties in the AME2003 data are all smaller than size of the symbols used to represent the data points.

It should be noted that in figure 2 we have only considered one nuclide out of many isobars for each mass number $A$. We shall now analytically prove that the nuclides we have considered, for various atomic mass numbers, have the highest binding energy per nucleon compared to its isobars. Consider the following semi-empirical mass formula.

$$M_{atom}(A,Z) \approx Zm(^{1}\text{H}) + Nm_n - \frac{BE(A,Z)}{c^2} \qquad (10)$$

Here, $c$ represents the speed of light and $M_{atom}(A,Z)$ represents the atomic mass of the isobar (represented symbolically as $^{A}$X), with atomic number $Z$ and mass number $A$. $m(^{1}\text{H})$ represents mass of the hydrogen atom and $m_n$ is the mass of the neutron. The nuclear binding energy $BE(A,Z)$ is assumed to take the form shown in equation (9). The atomic masses $M_{atom}$ of various isobars are plotted against their atomic numbers $Z$ in figure 3. We have considered only even–even nuclei in figure 3 since they have higher binding energy per compared to odd–odd nuclei. For odd–odd nuclei the parabola has exactly the same form as shown in figure (3) for even–even nuclei, but will be shifted up due to the parity term. For even–even nuclei, the unstable isobars approach stability by converting a neutron into a proton or a proton into a neutron by shifting between these two parabolae.

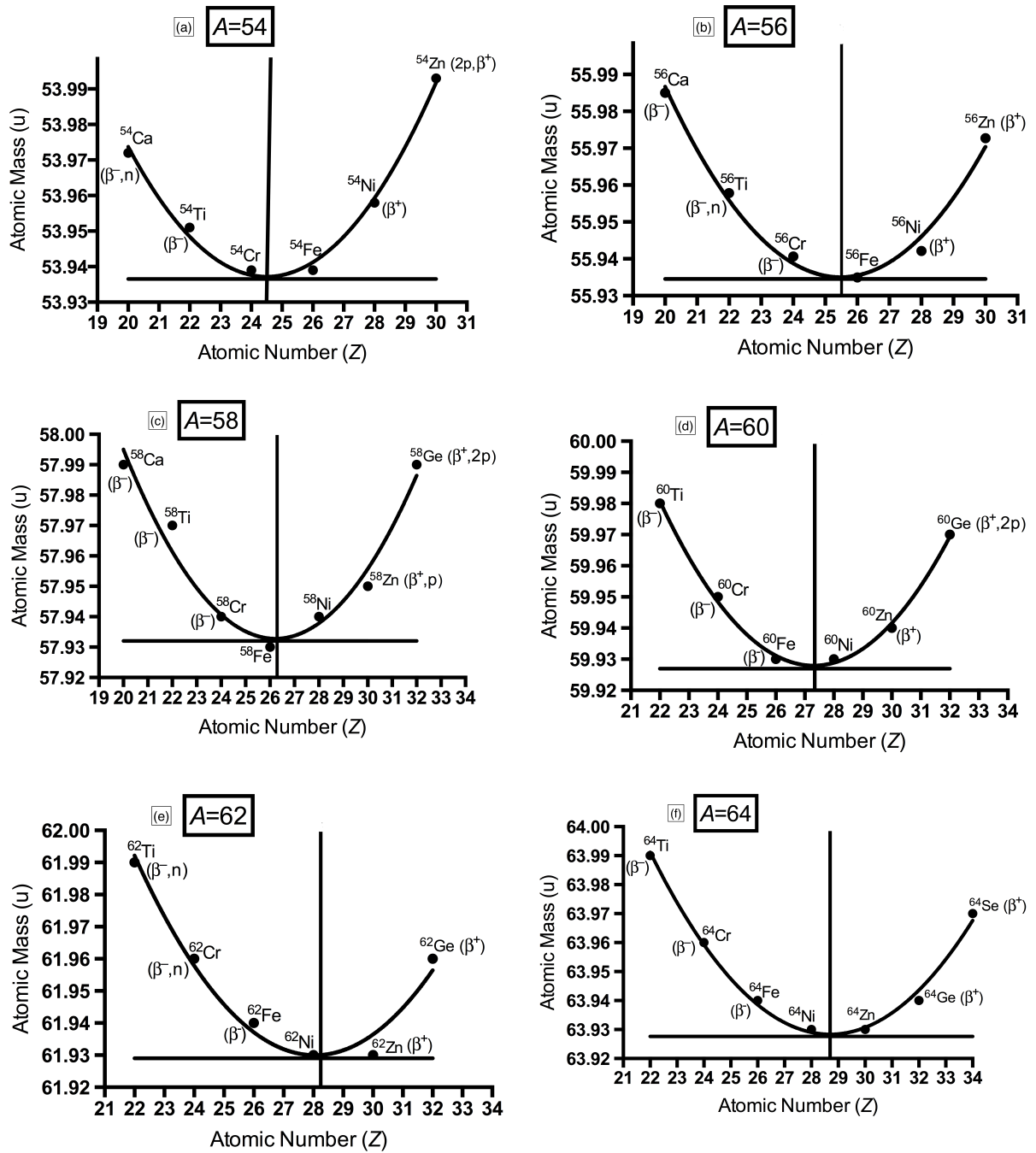

**Figure 3:** Mass chains for all the mass numbers $A$ considered in figure 1. The decay modes of each nuclide are shown in parenthesis. The atomic masses, measured in standard atomic mass units (u), are taken from AME2003. The solid smooth lines are least-square fitted parabolae using the non-linear quadratic regression model in GraphPad's Prism 7.0c, since the binding energy function is quadratic in $Z$. The horizontal and vertical lines are drawn to show the atomic number $Z$ that minimises the atomic mass, for each value of $A$. The uncertainties are all smaller than the size of the symbols used to represent the data points.

It can be seen in figure 3 that the stable isobars are nuclides that have the lowest mass in mass chain curves. If we use the approximation that $m(^1\text{H}) \approx m_n = m$, equation (10) can be written as

$$M_{atom}(A,Z) \approx A\left[m - \frac{BE(A,Z)}{Ac^2}\right] \tag{11}$$

From equation (11), it is clear that we can minimise the atomic mass of $^A\text{X}$ by maximising $BE/Ac^2$. Hence, we can alternatively say that the stable nuclides in figure (3) correspond to highest binding energy per nucleon. Therefore, for stable isobars, we have

$$\frac{\partial BE(A,Z)}{\partial Z} = 0 \tag{12}$$

Using the binding energy form given by equation (9), we have

$$\frac{-a_c(2Z-1)}{A^{4/3}} + \frac{4a_a(A-2Z)}{A^2} = 0 \tag{13}$$

Upon rearranging the terms, we get

$$Z = \frac{A}{2}\left[\frac{1 + \frac{a_c}{4a_a}A^{-1/3}}{1 + \frac{a_c}{4a_a}A^{2/3}}\right] \tag{14}$$

The atomic numbers of various nuclides with highest mean binding energy per nucleon for different mass numbers for $54 \leq A \leq 60$ are shown in table 3. The minimum values from graphs shown in figure 3 are also given in table 3 for comparison.

**Table 3:** The nuclei with the highest mean binding energy per nucleon, and correspondingly lowest mass, for each value of atomic mass number (*A*) from theoretical considerations using equation (14) and minimum values from least-square fits shown in figures 3. The values from least-square fits are taken from intersection of the x-axis and the vertical lines shown in figure 3 using GraphPad's Prism 7.0c.

| Mass number (*A*) | Theoretical calculation (*Z*) | Least-square fits (*Z*) |
|---|---|---|
| 54 | 24.39 | 24.5 |
| 56 | 25.24 | 25.4 |
| 58 | 26.08 | 26.1 |
| 60 | 26.91 | 27.2 |
| 62 | 27.74 | 28.1 |
| 64 | 28.57 | 28.8 |

It is evident from table 3 that the theoretical calculations of atomic numbers (*Z*) and minimum values of *Z* obtained from least-square fits of the experimentally available atomic masses are approximately equal. Nevertheless, there are slight differences in these values because we have considered only few terms in the binding energy formula. These sets of $(A, Z)$ shown in table 3 have the highest mean binding energy per nucleon and are shown in figures 1 and 2. Alternatively, the isobars closest to the vertical lines in figure 3 have highest mean binding energy per nucleon and, hence, are considered in figures 1 and 2.

### III. The local extrema of the binding energy function

As outlined in the section II, we shall consider the following form of the binding energy per nucleon for the even–even nuclei

$$\frac{BE(A,Z)}{A} = a_v - \frac{a_s}{A^{1/3}} - a_c \frac{Z(Z-1)}{A^{4/3}} - a_a \frac{(N-Z)^2}{A^2} + \frac{a_p}{A^{3/2}} \quad (15)$$

The values for the empirical constants that are found to be the best fit to AME2003 data are $a_v = 15.7827$ MeV, $a_s = 17.9042$ MeV, $a_c = 0.724040$ MeV, $a_a = 23.7193$ MeV and $a_p = 11.0000$ MeV [16]. We have assumed the significance of six digits for these coefficients to be consistent throughout our analysis. We only consider the even–*Z* and even–*N* nuclei, since, due to the parity term, they have larger mean binding energy per nucleon compared to other combinations of *Z* and *N*. In order to find the maximum value of *BE/A*, we need to maximize *BE/A* across the whole range of atomic nuclei. We can do this by maximizing *BE/A* subject to the condition that *A=N+Z*. Thus, we have the following Lagrangian problem:

$$Maximise: a_v - \frac{a_s}{A^{1/3}} - a_c \frac{Z(Z-1)}{A^{4/3}} - a_a \frac{(N-Z)^2}{A^2} + \frac{a_p}{A^{3/2}}$$

$$Subject\ to: A - N - Z = 0$$

The Lagrange function *L* for the defined problem, with a multiplier $\lambda$, then is

$$L = a_v - \frac{a_s}{A^{1/3}} - a_c \frac{Z(Z-1)}{A^{4/3}} - a_a \frac{(N-Z)^2}{A^2} + \frac{a_p}{A^{3/2}} + \lambda(A - N - Z) \quad (16)$$

Following the standard procedure, we set $\frac{\partial L}{\partial A} = \frac{\partial L}{\partial N} = \frac{\partial L}{\partial Z} = 0$. We then have

$$\frac{\partial L}{\partial A} = \frac{a_s}{3A^{4/3}} + \frac{4a_c Z(Z-1)}{3A^{7/3}} + \frac{2a_a(N-Z)^2}{A^3} - \frac{3a_p}{2A^{5/2}} + \lambda = 0 \quad (17)$$

$$\frac{\partial L}{\partial N} = \frac{2a_a(N-Z)}{A^2} + \lambda = 0 \quad (18)$$

$$\frac{\partial L}{\partial Z} = \frac{a_c(2Z-1)}{A^{4/3}} - \frac{2a_a(N-Z)}{A^2} + \lambda = 0 \qquad (19)$$

Next, eliminating the multiplier $\lambda$ in equations (17), (18) and (19), and using $N = A - Z$, we have the following two equations.

$$\frac{a_s}{3A^{4/3}} + \frac{4a_c Z(Z-1)}{3A^{7/3}} + \frac{2a_a(A-2Z)^2}{A^3} - \frac{3a_p}{2A^{5/2}} - \frac{2a_a(A-2Z)}{A^2} = 0 \qquad (20)$$

$$\frac{4a_a}{a_c}(A-2Z) = (2Z-1)A^{2/3} \qquad (21)$$

Upon rearranging equation (21), we see that we arrive at a familiar result [23]:

$$Z = \frac{A}{2}\left[\frac{1 + \frac{a_c}{4a_a}A^{-1/3}}{1 + \frac{a_c}{4a_a}A^{2/3}}\right] \qquad (22)$$

Substituting $Z$ given by equation (22) in equation (20), we get the following 15$^{th}$ degree polynomial equation.

$$\left.\begin{array}{l}\dfrac{a_s a_c^2}{3}k^{15} - \dfrac{8a_c a_a^2}{3}k^{13} + \left[\dfrac{8a_a a_s a_c}{3} - 2a_a a_c^2\right]k^{11} - \dfrac{a_c^3}{3}k^9 - \dfrac{3a_p a_c^2}{2}k^8 \\ + \left[\dfrac{16a_s a_a^2}{3} - \dfrac{8a_c a_a^2}{3}\right]k^7 - \dfrac{2a_a a_c^2}{3}k^5 - 12a_a a_c a_p k^4 - 24a_p a_a^2 = 0\end{array}\right\} \qquad (23)$$

For the sake of simplicity, in equation (23), we have defined $k^6 = A$. Upon substitution of the empirical constants in equation (23), we have

$$\left.\begin{array}{l}3.12866k^{15} - 1086.26k^{13} + 795.082k^{11} - 0.126522k^9 - 8.64986k^8 \\ +52636.38k^7 - 8.28964k^5 - 2266.93k^4 - 148527.8 = 0\end{array}\right\} \qquad (24)$$

The equation (24) is a polynomial equation in $k$ and it can be solved using numerical methods such as Newton's method or a standard computational knowledge engine such as Wolfram Alpha [24]. Using Wolfram Alpha, we note that equation (24) has 5 real solutions and they are $k \approx \pm 18.6135$, $k \approx -1.98954$, $k \approx 1.16853$ and $k \approx 1.97137$. The solutions, in $A \approx 60$ region, are represented by $k \approx -1.98954$ and $k \approx 1.97137$, and they correspond to $A \approx 62.0178$ and $A \approx 58.6960$, respectively. These represent the local extrema of the semi-empirical mass formula. The corresponding values of the fractional atomic numbers $Z$ can be found by substituting theses values of $A$ in equation (22) and upon substitution of these values of $A$, we get $Z \approx 27.7506$ for $A \approx 62.0178$ and $Z \approx 26.3908$ for $A \approx 58.6960$. Hence, the highest mean binding energy per nucleon, according to the liquid drop model, should be found near $A=62$, $Z=28$ ($^{62}$Ni) or $A=58$, $Z=26$ ($^{58}$Fe). The nuclide with the highest mean binding energy per nucleon (global maximum) can be found using the formula

$$\frac{BE(A,Z)}{A} \approx \frac{1}{A}\left(Zm(^1\text{H}) + Nm_n - M_{atom}(A,Z)\right)c^2 \qquad (25)$$

Using the tabulated values for the terms in the equation (25), it can be shown that $^{62}$Ni has the highest mean binding energy per nucleon [10]. Moreover, as mentioned before, $^{62}$Ni has 28 protons and they form a closed shell and is expected to have higher mean binding energy per nucleon than $^{58}$Fe.

## IV. The abundance of $^{62}$Ni compared to $^{56}$Fe

We have seen that our current nuclear model predicts that $^{62}$Ni is the most tightly bound nucleus. However, experimental data suggests that $^{56}$Fe is the sixth most abundant nuclide in the solar system [25]. In order to understand this difference, we must understand the nuclear burning processes that occur in the core of massive stars. It must also be noted that only a few reactions have non-negligible cross-sections during a certain phase (that depends on temperature and mass of the star) and the most important processes are Hydrogen, Helium, Carbon, Oxygen and Silicon burning. We shall now present a brief discussion of these processes under equilibrium conditions and concentrate more on the dominant end products of various burning stages. We shall begin by discussing the gravitational contraction of a huge interstellar molecular cloud that is not only particularly rich in $^1$H, $^4$He, but may also contain traces of other nuclides such as $^6$Li, $^{7,8}$Be and $^8$B. For a more detailed discussion of these processes, the reader is advised to refer to an excellent textbook by Jordi José [26] and the original papers by B$^2$FH [27] and Cameron [28].

*4.1. Hydrogen burning*

Since the initial configuration of the molecular cloud is hydrogen, it forms the first burning stage of the stellar evolution. Since the hydrogen nuclei have only one proton, the hydrogen burning stage requires relatively low temperatures (~$10^7$ K). In stellar plasmas, the essential reaction of the hydrogen burning stage is consumption of 4 $^1$H to form a $^4$He nucleus as shown in equation (26).

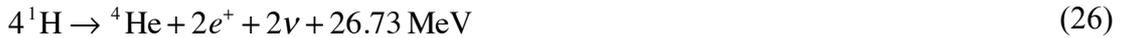

$$4\,^1\text{H} \rightarrow\ ^4\text{He} + 2e^+ + 2\nu + 26.73\,\text{MeV} \qquad (26)$$

It should be noted that the net energy released in equation (26) is calculated using the binding energy differences in $^4$He and four $^1$H nuclei, together with the energy released due to the annihilation of positrons ($e^+$) by electrons. A small fraction of this energy is carried away by neutrinos ($\nu$). The hydrogen burning reactions proceed through two dominant mechanisms in stellar plasmas: proton–proton chains and CNO cycles. We shall not go into the further details of these reactions, except to note that they differ in the number of neutrinos emitted and also in the final energy released.

*4.2. Helium burning*

The second stage of stellar evolution is burning of helium nuclei and this happens, after the hydrogen in the core is exhausted, for stars with mass at least 0.5 times the mass of the Sun ($M_0$). The core of the star contracts and this increases the temperature to $10^8$ K, at which point the helium burning starts. The helium burning occurs in two steps. The first step is formation of an unstable nuclide $^8$Be, which has a half-life of $10^{-16}$ seconds. Since a lot of helium nuclei fuse, an equilibrium in concentration of $^8$Be is established. The second step involves alpha-capture of $^8$Be nuclei resulting the formation of $^{12}$C nuclei. Hence, this reaction is also called the or triple-$\alpha$ reaction. The steps, along with the energy releases, are shown in equations (27).

$$\left.\begin{array}{l}^{4}\text{He} + {}^{4}\text{He} \rightarrow {}^{8}\text{Be} + \gamma - 91.8\text{ keV} \\ {}^{4}\text{Be} + {}^{4}\text{He} \rightarrow {}^{12}\text{C} + 7.37\text{ MeV}\end{array}\right\} \qquad (27)$$

The next dominant exothermic reaction that immediately follows after the creation of $^{12}$C nuclei results in formation of $^{16}$O nuclei.

$$^{12}\text{C} + {}^{4}\text{He} \rightarrow {}^{16}\text{O} + \gamma + 7.16\text{ MeV} \qquad (28)$$

Several other reactions may take place depending on the temperature of the stellar plasma. If the temperature is greater than 0.4 GK, $^{12}$C($\alpha,\gamma$)$^{16}$O is followed by $^{16}$O($\alpha,\gamma$)$^{20}$Ne and subsequently $^{20}$Ne($\alpha,\gamma$)$^{24}$Mg. A few more reactions, with varying cross-sections and probabilities, are possible in the presence of $^{14}$N, but we shall not go into the details. The dominant end product of the helium burning stage is $^{12}$C, which fuels the next stage of stellar evolution.

*4.3. Carbon burning*

When the mass of the star is greater than 7 $M_0$, carbon burning processes begin to establish an equilibrium condition in star at temperatures greater than 0.6 GK. The dominant exothermic reactions in carbon burning stage are

$$\left.\begin{array}{l}^{12}\text{C} + {}^{12}\text{C} \rightarrow {}^{23}\text{Na} + {}^{1}\text{H} + 2.24\text{ MeV} \\ {}^{12}\text{C} + {}^{12}\text{C} \rightarrow {}^{20}\text{Ne} + {}^{4}\text{He} + 4.61\text{ MeV}\end{array}\right\} \qquad (29)$$

At the temperatures at which these reactions happen, all the emitted protons and alpha particles get absorbed by different nuclides present in stellar plasma to form $^{20}$Ne, through reactions such as $^{12}$C $(p,\gamma)^{13}$N$(\beta^+)^{13}$C$(\alpha,n)^{16}$O$(\alpha,\gamma)^{20}$Ne. At the end of carbon burning phase, the stellar plasma mostly contains $^{16}$O, $^{20}$Ne and $^{23}$Na. This forms the fuel for the next stage of stellar evolution, Neon burning.

*4.4. Neon burning*

The neon burning in hydrostatic equilibrium phase occurs when the temperature of the stellar plasma ranges approximately from 1.2–1.8 GK. The dominant reaction at this temperature is the photodisintegration of $^{20}$Ne, since it exhibits a low alpha-separation energy of 4.73 MeV. Hence, the dominant reaction at this stage is

$$^{20}\text{Ne} + \gamma \rightarrow {}^{16}\text{O} + {}^{4}\text{He} - 4.73\,\text{MeV} \tag{30}$$

The alpha particles produced in this process are subsequently captured by the remaining $^{20}$Ne nuclides in stellar plasma to form $^{28}$Si. A dominant path toward $^{28}$Si can be represented as $^{20}\text{Ne}(\alpha,\gamma)^{24}\text{Mg}(\alpha,\gamma)^{28}\text{Si}$. The dominant nuclides at the end of neon burning therefore are $^{16}$O, $^{28}$Si and $^{24}$Mg, which forms the fuel for the next burning stage.

### 4.5. Oxygen burning

When the temperature of the stellar plasma increases to 1.5–2.7 GK, the oxygen nuclides start burning. When to oxygen nuclei fuse, they form a metastable $^{32}$S compound nuclide. It decays into its stable isotopes by emitting either an alpha particle of two protons. The two dominant exothermic reactions that occur at this stage of stellar evolution are

$$\left.\begin{array}{l}^{16}\text{O} + {}^{16}\text{O} \rightarrow {}^{28}\text{Si} + {}^{4}\text{He} + 9.59\,\text{MeV} \\ ^{16}\text{O} + {}^{16}\text{O} \rightarrow {}^{30}\text{Si} + 2\,{}^{1}\text{H} + 0.38\,\text{MeV}\end{array}\right\} \tag{31}$$

The various other exothermic reactions that can occur at this stage form $^{31}$P and $^{31}$S nuclides in the plasma, so that the stellar plasma after oxygen burning phase is dominated by $^{28}$Si, $^{32}$S and $^{31}$P.

### 4.6. Silicon burning

The dominant process that occurs in silicon burning phase is the photodisintegration of loosely bound nuclei in the stellar plasma. The temperature of the core should be greater than 2.8 GK for silicon burning phase to begin. At these temperatures, photons have enough energy to disintegrate nuclides in plasma. The reactions that typically occur are $^{32}\text{S}(\gamma,\alpha)^{28}\text{Si}$ and $^{32}\text{S}(\gamma,p)^{31}\text{P}$. The $^{31}$P nuclides formed by such reactions are converted to $^{28}$Si by a suite of secondary reactions such as $^{31}\text{P}(\gamma,p)^{30}\text{Si}(\gamma,2n)^{28}\text{Si}$. The alpha, neutron and proton separation energies for $^{28}$Si are 9.98, 17.2 and 11.6 MeV, respectively. Hence, at temperatures greater than 2 GK, $^{28}$Si nuclides in stellar plasma undergoes various $(\gamma,\alpha)$, $(\gamma,n)$ and $(\gamma,p)$ reactions that create a sea of alpha particles, neutrons and protons. While these photon-induced disintegration reactions tend to decrease the atomic mass number of constituents of the stellar plasma, the reverse reactions such as alpha-capture reactions tend to increase the mass number. This creates quasi-equilibrium clusters of nuclides around $^{28}$Si that extends up to A ~ 40 and Fe-group elements with A ≥ 50.

### 4.7. Nuclear Statistical Equilibrium

After the depletion of $^{28}$Si in the core, it begins to contract at temperatures approximately equal to 4 GK. All the nuclides in stellar equilibrium from $^{1}$H to Fe-group that have been created by various burning processes until now form a quasi-equilibrium cluster called Nuclear Statistical Equilibrium (NSE). The abundance of various nuclides in a NSE depends on various parameters such as the temperature, neutron excess and density of stellar plasma. It can be proved, using the Saha equation [29], that relatively low temperature of stellar plasma in NSE creates a distribution of abundances of nuclides that favour $^{56}$Ni. $^{56}$Ni is unstable to beta decay and decays into $^{56}$Fe, which, as we have seen, has a high value of the binding energy per nucleon. We have seen in our discussions that the stellar plasma is rich in alpha particles, neutrons and protons in silicon burning phase. As pointed out by R Shurtleff and E Derringh in their letter [10], there exists no reaction to bridge the gap from $^{56}$Fe to $^{62}$Ni with these particles. Hence, $^{56}$Fe is the end product of the nuclear burning processes because it is in close proximity to $^{62}$Ni, which is the most tightly bound nucleus and therefore, $^{56}$Fe has relatively higher abundance than $^{62}$Ni and $^{58}$Fe.

## V. Conclusions and implications for teaching

The analytical proof presented in this paper can be a useful tool to the instructors to introduce important concepts, while teaching binding energy of atomic nuclei. One of the solutions of equation (24), for instance, is $k \approx 1.16853$ and the corresponding value of fractional mass number is $A \approx 2.54588$. This resembles the $^4$He nuclide, which has a very high value of binding energy per nucleon. However, it should be noted that a solution closer to $A = 4$ was not obtained because G Royer and C Gautier have considered 1522 nuclei with $Z$ and $N > 7$ [22]. Nevertheless, our analysis predicts a $^4$He-like nuclide that must have an exceptionally high value of binding energy per nucleon. This analysis can also be used to explain why alpha-conjugate nuclides have higher abundance in the solar system. The other solution that was obtained was $k \approx \pm 18.6135$ and this corresponds to nuclide with approximately $A \approx 4.7 \times 10^7$ number of nucleons. We shall ignore this solution because an object with $4.7 \times 10^7$ number of nucleons has a large negative value of binding energy per nucleon and cannot form a stable system.

We can also explain why heavy stable nuclei with $A > 40$ have $N \neq Z$, by considering equation (22). The stable nuclei, for various combinations of atomic and mass numbers, have a high value of the mean binding energy per nucleon. Hence, equation (22) represents an approximate[2] condition to be satisfied for a nuclide to be stable. Upon substitution of value of the empirical constants, we have

$$Z \approx \frac{A}{2}\left[\frac{1+0.0076 A^{-1/3}}{1+0.0076 A^{2/3}}\right] \tag{32}$$

For relatively low values of $A$, from equation (32), we have $Z \approx A/2$ for stable nuclei, but for heavier stable nuclides, we have $Z < A/2$, consistent with the observed AME2016 data. The analysis presented in this paper can also be used by the instructors to introduce various Atomic Mass Evaluations and the corresponding different sets of coefficients of the Bethe-Weizsäcker and binding energy formulae.

---

[2] It should be noted that equation (22) becomes an exact condition for stability if we let the atomic number and mass number take on continuous fractional values, and if we consider the binding energy function given by equation (9). Since the atomic nuclides in nature cannot take on such fractional values, this represents an approximate condition that all stable nuclides strive to satisfy.

The subtle relationship between stability and abundance of a nuclide is often not emphasised. The brief discussion, presented in section IV, of the various nuclear burning processes that occur in the cores of massive stars can help to understand why $^{56}$Fe is more abundant than $^{62}$Ni. This can also help to eliminate the misconception that $^{56}$Fe is the most tightly bound nucleus. Without going into the complicated numerical analyses, the introduction of the Lagrange problem presented in this paper can provide a unique opportunity for the students to learn the application of it in nuclear physics, while teaching an important idea that a nuclide need not be abundant just because it has the highest mean binding energy per nucleon.

## VI.    Acknowledgements

I would like to thank the two anonymous referees for their comments and insights that helped in substantially improving the quality of the manuscript. I wish to thank Dr. Ross Springell and Ms. Aakriti Batra of the University of Bristol for going through the manuscript and verifying the tedious algebraic manipulations. I would also like to thank Mr. Mukund Srinath of Pennsylvania State University and Ms. Swagata Dutta and Ms. Medha Shekhar of Georgia Institute of Technology for their help with finding the references.

**References**

[1]   H. A. Bethe, "Supernovae", *Phy. Today* **43** (9), 24–27 (1990).

[2]   J. Michael Pearson, "On the belated discovery of fission", *Phy. Today* **68** (6), 40–45 (2015).

[3]   A Beiser, *Physics*, 4$^{th}$ ed. (Cummings, Menlo Park, 1986), p. 794.

[4]   Arthur Beiser, *Concepts of Modern Physics*, 6$^{th}$ ed. (McGraw-Hill, New York, 2003) p. 401 (See figure 11.12).

[5]   J. W. Kane and M. M. Sternheim, *Physics*, 3$^{rd}$ ed. (Wiley, New York, 1988), p. 753.

[6]   T. M. Corwin and D. G. Wachowiak, *The Universe, from Chaos to Consciousness* (Harcourt Brace Jovanovich, San Diego, 1989), p. 240.

[7]   Vern J. Ostdiek and Donald J. Bord, *Inquiry into physics*, 7$^{th}$ ed. (Brooks/Cole, Cengage learning, Boston, 2013), p. 446.

[8]   V. Trimble, "The origin and abundances of the chemical elements", *Rev. Mod. Phys.* **47**, 877-976, p. 935 (1975).

[9]   M. P. Fewell, "The atomic nuclide with the highest mean binding energy", *Am. J. Phys.* **63**, 653 (1995).

[10]  R Shurtleff and E. Derringh, "The most tightly bound nucleus", *Am. J. Phys.* **57**, 552, (1989).

[11]  W. E. Burcham, *Nuclear Physics, an introduction*, 2$^{nd}$ ed. (Longman Group Limited, London, 1973), p. 264 (See point (d)).

[12] W. E. Burcham and M Jobes, *Nuclear and Particle Physics* (Pearson Education Limited, Essex, 1995), p. 96 (See point (c)).

[13] Paul A. Tipler and Gene Mosca, *Physics: For scientists and Engineers,* 6th ed. (W. H. Freeman and Company, New York, 2008), fig. 40–3, ch. 40, p. 1362.

[14] Meng Wang *et al*, "The AME2016 atomic mass evaluation (II). Tables, graphs and references", *Chinese Phys. C* **41**, 030003 (2017).

[15] H. A. Bethe and R. F. Bacher, "Nuclear Physics A. Stationary States of Nuclei", *Rev. Mod. Phys.* **8**, 82 (1936).

[16] G Royer and B Remaud, "Fission processes through compact and creviced shapes", *J. Phys. G: Nucl. Phys.* **26**, 1149 (1984).

[17] G Royer, "Alpha emission and spontaneous fission through quasi-molecular shapes", *J. Phys. G: Nucl. Phys.* **10**, 1057 (2000).

[18] Michael W. Kirson, "Mutual influence of terms in a semi-empirical mass formula", *Nucl. Phys. A* **798**, p. 29–60 (2008).

[19] M. Wang *et al*, "The AME2012 atomic mass evaluation", *Chinese Phys. C* **36** 1603 (2012).

[20] G. Audi, A. H. Wapstra, and C. Thibault, "The AME2003 atomic mass evaluation: Tables, graphs and references", *Nucl. Phys. A* **729**, 337 (2003).

[21] G. Royer and A. Subercaze, "Coefficients of different macro–microscopic mass formulae from the AME2012 atomic mass evaluation", *Nucl. Phys. A* **917**, p. 1–14 (2013).

[22] G. Royer and C. Gautier, "Coefficients and terms of the liquid drop model and mass formula", *Phys. Rev. C.* **73**, 067302 (2006).

[23] Kenneth S. Krane, *Introductory Nuclear Physics* (John Wiley and Sons, Inc., New York, 1988), p. 70 (See eq. (3.31)).

[24] Wolfram Alpha: https://www.wolframalpha.com

[25] Carl J. Hansen *et al. Stellar Interiors,* 2nd ed. (Springer New York, 1994), p. 79. (See fig. 2.19).

[26] Jordi José, *Stellar Explosions: Hydrodynamics and Nucleosynthesis* (CRC Press, Boca Raton, 2016), p. 93.

[27] E. Margaret Burbidge, G. R. Burbidge, William A. Fowler, and F. Hoyle, "Synthesis of the Elements in Stars*", Rev. Mod. Phys.* **29**, 547 (1957).

[28] A. G. W. Cameron, *Stellar Evolution, Nuclear Astrophysics, and Nucleogenesis,* 2nd ed. (Dover Publications, Mineola (New York), 2013).

[29] See p. 112 of Ref [26].